\documentclass[letterpaper]{article} 
\usepackage{aaai25}  
\usepackage{times}  
\usepackage{helvet}  
\usepackage{courier}  
\usepackage[hyphens]{url}  
\usepackage{graphicx} 
\usepackage{multirow}
\usepackage{natbib}  
\usepackage{caption} 
\usepackage{makecell}
\usepackage{algorithm}
\usepackage{algorithmic}

\usepackage{newfloat}
\usepackage{listings}
\DeclareCaptionStyle{ruled}{labelfont=normalfont,labelsep=colon,strut=off} 
\floatstyle{ruled}
\newfloat{listing}{tb}{lst}{}
\floatname{listing}{Listing}
\title{Block-Based Multi-Scale Image Rescaling}
\author{
    Jian Li and Siwang Zhou\thanks{Corresponding author.}\\
}
\affiliations{
    Hunan University, China
    
    \{lijian543, swzhou\}@hnu.edu.cn

}

\usepackage{bibentry}

\begin{document}

\maketitle

\begin{abstract}

Image rescaling (IR) seeks to determine the optimal low-resolution (LR) representation of a high-resolution (HR) image to reconstruct a high-quality super-resolution (SR) image. Typically, HR images with resolutions exceeding 2K possess rich information that is unevenly distributed across the image. Traditional image rescaling methods often fall short because they focus solely on the overall scaling rate, ignoring the varying amounts of information in different parts of the image. To address this limitation, we propose a Block-Based Multi-Scale Image Rescaling Framework (BBMR), tailored for IR tasks involving HR images of 2K resolution and higher. BBMR consists of two main components: the Downscaling Module and the Upscaling Module. In the Downscaling Module, the HR image is segmented into sub-blocks of equal size, with each sub-block receiving a dynamically allocated scaling rate while maintaining a constant overall scaling rate. For the Upscaling Module, we introduce the Joint Super-Resolution method (JointSR), which performs SR on these sub-blocks with varying scaling rates and effectively eliminates blocking artifacts. Experimental results demonstrate that BBMR significantly enhances the SR image quality on the of 2K and 4K test dataset compared to initial network image rescaling methods.

\end{abstract}

%

\section{Introduction}
With the growing demand in practical applications, the technology of image rescaling (IR) has rapidly advanced. Specifically, it can significantly optimize storage utilization \cite{image_compression_downsampling} and minimize the bandwidth required for images and videos transmission \cite{nas, yeo2020nemo, yu2023bisr}. The availability of high-resolution displays and advancements in camera equipment have led to the widespread usage of images and videos with 2K resolution or higher. However, as image resolution increases, there is a significant exponential growth in the demand for storage space and network transmission bandwidth. Therefore, the development of IR techniques is crucial for effectively storing and transmitting high-resolution images and videos.

\begin{figure}[!ht]
\setlength{\abovecaptionskip}{0.cm}
\begin{center}
\centering
\includegraphics[width=0.47\textwidth]{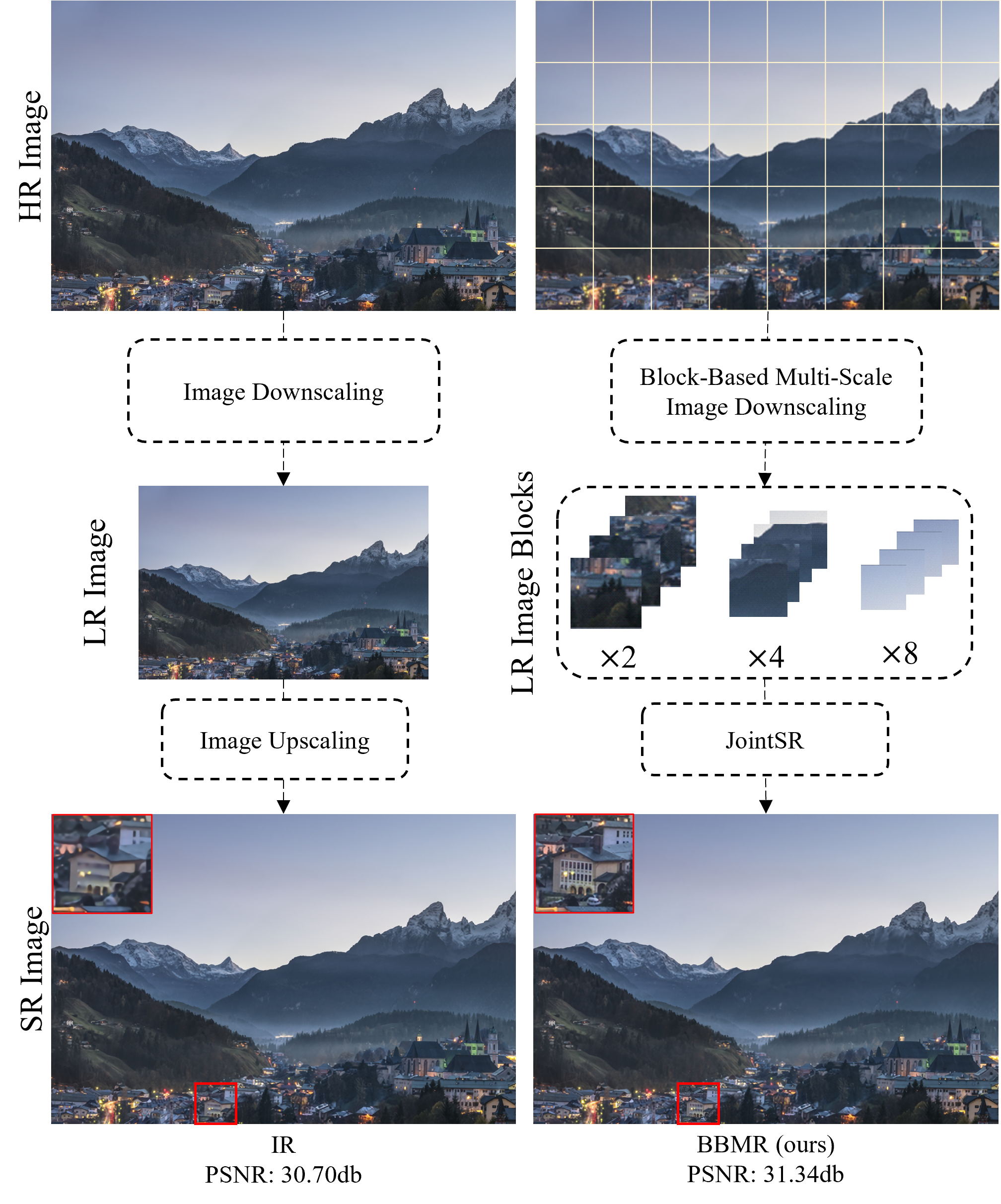}

\end{center}
   \caption{Comparison of traditional image rescaling methods (left) and our block-based multi-scale image rescaling framework (right) at ×4 scaling rate. Both methods use bicubic interpolation downscaling and OmniSR model upscaling. The improvement in the image quality of the red box is due to the smaller scaling rate.}
\label{fig:example}
\vspace{-0.2cm}
\end{figure}

Typically, IR methods \cite{learned_imgdown, HCD, invertible} demonstrate superior super-resolution (SR) image reconstruction performance compared to image super-resolution models \cite{wang2023omni, HAT, DAT} trained with bicubic downscaling. This superiority is attributed to IR’s focus on designing well-crafted downscaling methods, which produce low-resolution (LR) images more suitable for SR reconstruction. In practice, from a holistic perspective, these methods emphasize overall image scaling. However, \cite{kong2021classsr} suggests that the difficulty of super-resolution varies across different regions of an image, indicating that different parts contain varying amounts of information. Therefore, considering only the overall scaling rate limits the reconstruction performance of IR methods.

In this paper, we consider an application scenario for image rescaling where LR images are utilized solely to save storage space and reduce transmission bandwidth without the need for visualization. Inspired by \cite{zhou2020multi}, which improves overall reconstruction quality by dynamically adjusting the blocks' compression ratio, we adopt a similar approach for image rescaling. Specifically, we select different downscaling factors for different parts of the image, enhancing the overall super-resolution quality.

In this work, we introduce the BBMR framework for achieving multi-scale scaling, which comprises two components: the Downscaling Module and the Upscaling Module. Within the Client/Server Architecture, the Downscaling Module is typically managed on the server side, rendering the computational cost negligible. In this module, the HR image is initially divided into 128×128 sub-blocks. Subsequently, we dynamically allocate the scaling rate for each sub-block while maintaining a consistent overall scaling rate, utilizing the assistance of the super-resolution model from the Upscaling Module during this process. For instance, regions of the image containing sky are assigned a higher scaling rate, whereas areas with buildings are allocated a lower scaling rate. As illustrated in Fig. \ref{fig:example}, our block-based multi-scale downscaling method (right) contrasts with traditional downscaling methods (left). In the Upscaling Module, we utilize a joint super-resolution method (JointSR) to perform sub-block super-resolution and eliminate blocking artifacts. This method integrates a deblocking branch to address blocking effects at the feature level. Furthermore, we observe that deep learning-based SR models share certain similarities. To minimize server-side computation, a lightweight super-resolution model can be employed to assist in scaling rate allocation with minimal degradation of SR image quality.

Since what we propose is a general image rescaling framework that can be applied to various image rescaling methods, we used two different SR networks, OmniSR \cite{wang2023omni} and CRAFT \cite{CRAFT}, in our experiments. Additionally, we employed the neural network downscaling method described in \cite{global} and bicubic interpolation to validate our BBMR framework.

Overall, our contributions can be summarized in three key aspects: 
\begin{itemize}
    \item We proposed the block-based multi-scale downscaling strategy, which can more reasonably allocate scaling rates for different parts of the image in the Downscaling Module, thereby significantly improving the image reconstruction quality.
    
    \item We propose the JointSR method, which adopts a transfer learning strategy and incorporates LR blocks of different resolutions for training. Additionally, it introduces a deblock branch to eliminate blocking artifacts at the feature level.
    
    \item We find that it can use a very lightweight super-resolution model to assist in scaling rate allocation in the Downscaling Module, which can greatly reduce the computational complexity of the Downscaling Module.
\end{itemize}

\label{sec:intro}

\section{Related work}
\label{sec:formatting}
\subsection{Single Image Super-Resolution}
With the rapid development of deep learning, significant progress has been made in single image super-resolution. SRResNet \cite{srresnet} utilized residual connections to preserve details from previous layers. Recent approaches, such as \cite{swinir, HAT, DAT}, employ global receptive field attention mechanisms to further enhance SR quality. In addition, advancements in diffusion models \cite{gao2023implicit_diffusion, yue2024resshift_diffusion} have been explored in the image SR process to generate clearer super-resolved images.

\subsection{Block-Base Super-Resolution}
Block-based SR methods have also been extensively studied. \cite{STDO} proposes a video block over-fitting super-resolution method to achieve real-time and high-quality video super-resolution. Other methods, such as \cite{kong2021classsr, wang2024camixersr_block, zhang2024transformer_block}, adopt different types of super-resolution models according to the difficulty of super-resolving image sub-blocks to improve the overall super-resolution speed. \cite{luo2024adaformer} leverages the block characteristics of transformers, different categories of image sub-blocks to exit early in various Transformer layers to improve calculation speed. Unlike the above methods, which mainly focus on improving the super-resolution speed of the Upscaling Module, our proposed BBMR framework implements block multi-scaling in the Downscaling Module to enhance the image super-resolution quality of the Upscaling Module.

\subsection{Image Rescaling}
Image rescaling (IR) and image super-resolution (SR) are distinct tasks. IR involves downscaling and upscaling, while SR only focuses on upscaling. In IR, a ground-truth HR image is downscaled for storage and transmission and recovered when necessary. Downscaling, which generates an LR version of an HR image, is the inverse of SR. Bicubic interpolation \cite{BICUBIC} is the most common downscaling method. Recently, more and more work has been treating downscaling and upscaling as a unified process. \cite{learned_imgdown} proposed a learning-based image reduction method using a Content-adaptive Resampler, which effectively improves SR reconstruction. \cite{task-aware} introduced a technique called task-aware image reduction to support SR tasks. Moreover, \cite{invertible} presented a reversible network that models bidirectional degradation and recovery from a new perspective. These works focus on the overall scaling of the image, but in many scenarios, LR images are only used as storage and transmission media \cite{nas, yu2023bisr}. Therefore, we propose our BBMR framework, whose LR images are stored as sub-blocks with different resolutions, and can be applied to most of the IR and SR methods.


\begin{figure*}[htbp]
\begin{center}
\includegraphics[width=0.95\textwidth]{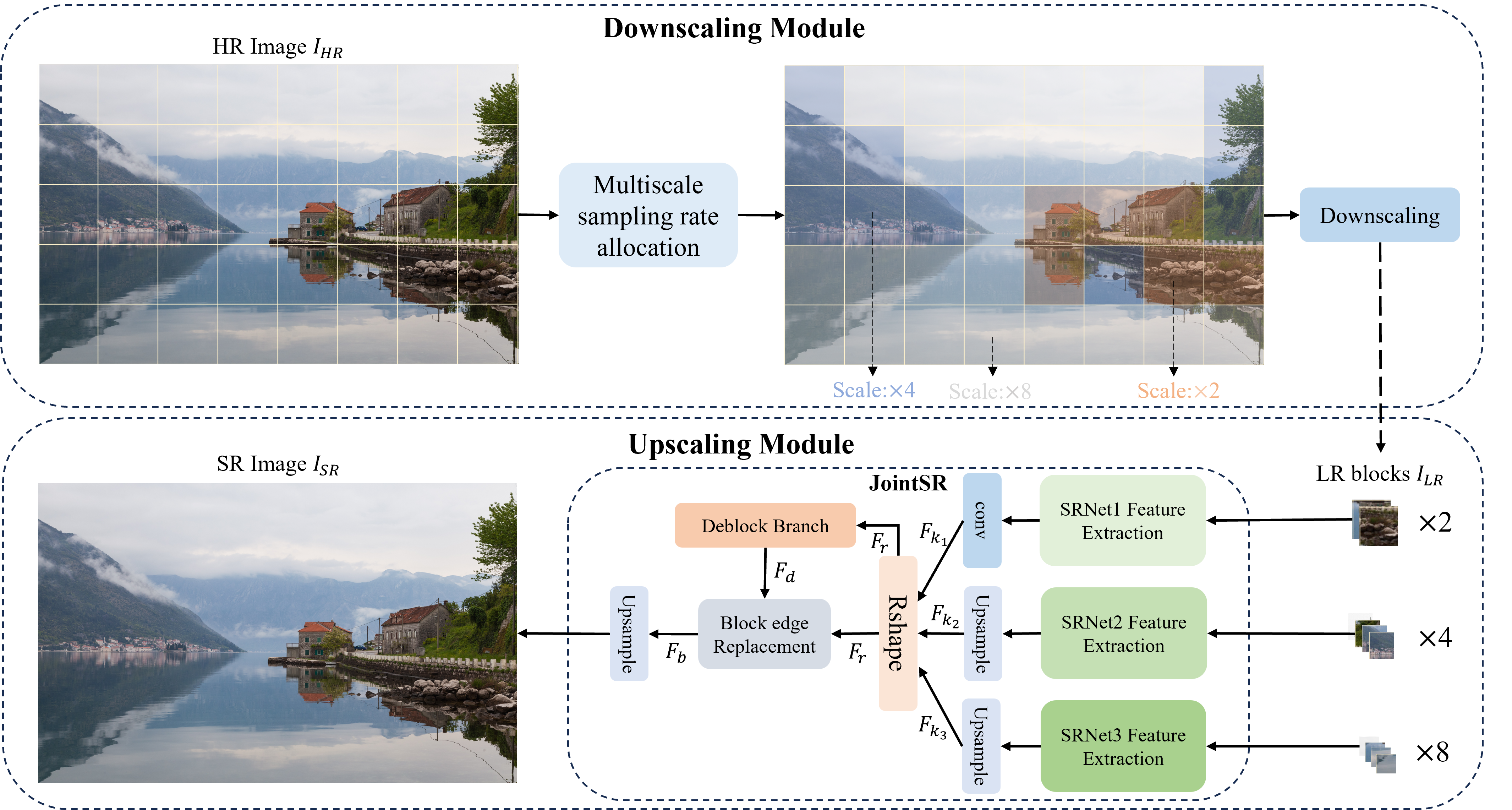}
\end{center}
   \caption{The overall framework structure of the proposed BBMR when the scaling rate is ×2, ×4 and ×8. Downscaling Module: Outputs LR image sub-blocks with three different resolutions through block-based multi-scale downscaling strategy. Upscaling Module: Generates high-quality SR images using the JointSR method. Upsample contains a convolution layer and a pixelshuffle module.}
\label{fig:struct}
\end{figure*}

\section{Methods}
The overall architecture of our method is illustrated in Fig. \ref{fig:struct}, with the Block-Based Multi-Scale Image Rescaling Framework (BBMR) comprising two main components: the Downscaling Module and the Upscaling Module. The Downscaling Module divides the image into blocks and dynamically allocates the scaling rate for each block. The Upscaling Module then takes these blocks with varying scaling rates and processes them through our JointSR method, effectively performing super-resolution on the image blocks and removing blocking artifacts.

\subsection{Observation}
In order to evaluate the difficulty of super-resolution for image sub-blocks with different resolutions, we averaged the HR images of the DIV2K validation dataset into 128×128 sub-blocks and validated them using BICUBIC downscaling and OmniSR upscaling methods. As shown in Tab. \ref{tab:PSNR_of_MS}, we found that approximately 20\% of sub-blocks experienced only a 1.06 dB decrease when downscaled by ×8 compared to ×4, which we term simple blocks. About 4.7\% of the sub-blocks showed an 8.21 dB improvement when downscaled by ×2 compared to ×4, which we term hard blocks. The remaining 75.3\% showed no significant difference in PSNR changes between downscaling by ×2 and ×8 compared to ×4, which we term medium blocks. This finding allows us to trade a small quality loss in simple sub-blocks for a significant quality gain in hard sub-blocks while maintaining the overall scaling rate.

\begin{table}[h]
\renewcommand\arraystretch{1.0}
    \centering
    \begin{tabular}{c|c|c|c} \hline 
         \makebox[0.08\textwidth][c]{Scales}&  \makebox[0.08\textwidth][c]{simple}&  \makebox[0.08\textwidth][c]{medium}& \makebox[0.08\textwidth][c]{hard}
\\ \hline 
         ×2&  -&  37.24& 39.13
\\ 
         ×4&  39.02&  31.62& 30.92
\\ 
         ×8&  37.96&  27.34& -
\\ \hline
    \end{tabular}
    \caption{PSNR at different super-resolution scales for different categories of image blocks in the DIV2K validation dataset. The tests use BICUBIC downscaling and OmniSR upscaling}
    \label{tab:PSNR_of_MS}
\vspace{-0.2cm}
\end{table}

\subsection{Block-Based Multi-Scale Downscaling Strategy}
In Downscaling Module in order to realize adaptive allocation of image sub-block sample rate, we designed a sample rate allocation algorithm. The method first determines three downscaling factors \(k_1,k_2,k_3 (k_1<k_2<k_3)\) and takes \(k_2\) as the overall scaling rate. To ensure that the overall scaling rate is maintained at \(k_2\), it is necessary to determine the values of $a$ and $c$. The ratio $a:c$ represents the proportion of image blocks with scaling factors \(k_1\) and \(k_3\). $a$ and $c$ must satisfy the following formula.
\begin{equation}
\{a(\frac{hw}{k_1^2})+c(\frac{hw}{k_3^2})\} \bmod {k_2}^2 = 0
\label{1}
\end{equation}


Where h and w is the height and width of the image blocks. The input HR image \(X^{HR}\) is divided into $N$ sub-blocks \(\left\{x_i^{HR}\right\}_{i=1}^N\) of size $(h, w)$, and then all the sub-blocks are downscaled into LR sub-blocks of three resolutions respectively, the process are as follows: 
\begin{equation}
\large
\{x_{i,s}^{LR}\}_{s=1}^3=\{f_{Down}^s\left(x_i^{HR}\right)\}_{s=1}^3
\label{1}
\end{equation}

Where $s=1,2,3$ correspond to \(k_1, k_2, k_3\), respectively. Then all the LR sub-blocks are super-resolved into 128×128 resolution SR blocks by the same SR method as the Upscaling Module, which is indicated below:
\begin{equation}
\large
\{x_{i,s}^{SR}\}_{s=1}^3=\{f_{up}^s\left(x_{i,s}^{LR})\right\}_{s=1}^3
\label{2}
\end{equation}

Calculate the PSNR values for all the SR and HR sub-blocks \(\{P_{i,s}\}_{i=1}^N=\{f_{PSNR}(x_{i}^{HR},x_{i,s}^{SR})\}_{i=1}^N\). Calculate the difference between all \(s=1\)and \(s=2\) sub-blocks in ascending order. Then, calculate the difference between all \(s=2\) and \(s=3\) sub-blocks in descending order sort. The process are as follows: 
\begin{equation}
\{{arr}_k^{earn}\}_{k=1}^N=sort(\{\ P_{i,1}-P_{i,2}\}_{i=1}^N,\ asc)
\label{1}
\end{equation}
\begin{equation}
\{{arr}_k^{pay}\}_{k=1}^N=sort(\{\ P_{i,2}-P_{i,3}\}_{i=1}^N,\ desc)
\label{1}
\end{equation}

The specific sample rate allocation algorithm is shown in Alg. \ref{sample rate allocate}.

\begin{algorithm}[ht]
    \renewcommand{\algorithmicrequire}{\textbf{Input:}}
    \renewcommand{\algorithmicensure}{\textbf{Output:}}
	\caption{Block Scaling Rate Allocation Algorithm}
        \label{sample rate allocate} 
        \begin{algorithmic}[1]
    	\REQUIRE{${arr}^{pay}$, ${arr}^{earn}$, $block\_max_{k_1}$, $a$, $c$, $t$, $N$} 
    	\ENSURE{${blocks}_{k_1}$, ${blocks}_{k_2}$, ${blocks}_{k_3}$}
            
            \FOR{$i = 1$ : $block\_max_{k_1}$}
                \STATE $earn\leftarrow Select({arr}^{earn}, i*a, a)$
                \STATE $pay\leftarrow Select({arr}^{pay}, i*c, c)$
                \IF{$pay+t > earn$}
                    \STATE $break$
                \ELSE
                    \STATE $Append(blocksk1 , {arr}^{earn}, i*a, a)$
                    \STATE $Append(blocksk3 , {arr}^{pay}, i*c, c)$
                \ENDIF
        \ENDFOR
        \STATE ${blocks}_{k_2}\leftarrow$ $FindB\_k2(N, {blocks}_{k_1}, {blocks}_{k_3})$\;
	\end{algorithmic}  
\end{algorithm}

$block\_max_{k_1}$is the maximum number of $k_1$ scaling factor blocks that can be selected. ${arr}^{pay}$ and ${arr}^{earn}$ represent $\{{arr}_k^{pay}\}_{k=1}^N$ and $\{{arr}_k^{earn}\}_{k=1}^N$ respectively. $t$ is a constant, set according to the super-resolution model and the image size. ${blocks}_{k_1}$, ${blocks}_{k_2}$, ${blocks}_{k_3}$ hold the sub-block numbers corresponding to the three scaling rates. $Select({arr}^{earn}, i*a, a)$ means selecting $a$ elements from ${arr}^{earn}$ starting at position $i*a$ and returning the sum of those $a$ elements. \(Append(blocksk1 , {arr}^{earn}, i*a, a)\) means selecting $a$ elements from ${arr}^{earn}$ starting at position $i*a$ to append at the end of blocksk1. $FindB\_k2$ means to return all blocks in $N$ that is not in ${blocks}_{k_1}$ and ${blocks}_{k_3}$.

\subsection{Joint Super-Resolution Method}
When stitching together super-resolved image blocks, block artifacts are inevitable. Typically, neural networks address block artifacts by processing the generated SR image, which significantly increases the computational load due to the input being in high resolution. To tackle this issue, we propose the joint super-resolution (JointSR) method, which performs deblocking while the features are still in the LR stage. As illustrated in Fig. \ref{fig:struct}, the SRNet Feature Extraction layer is derived from a pre-trained super-resolution model, where the final upscaling layer is removed. This part can be adapted depending on the super-resolution model used. Initially, LR image sub-blocks with different resolutions ${blocks}_{k_1}$, ${blocks}_{k_2}$, ${blocks}_{k_3}$ pass through the SRNet Feature Extraction and Upsample modules sequentially to obtain feature blocks of the same size ${F}_{k_1}$, ${F}_{k_2}$, ${F}_{k_3}$. 

\begin{figure}[htbp]
\begin{center}
\includegraphics[width=0.5\textwidth]{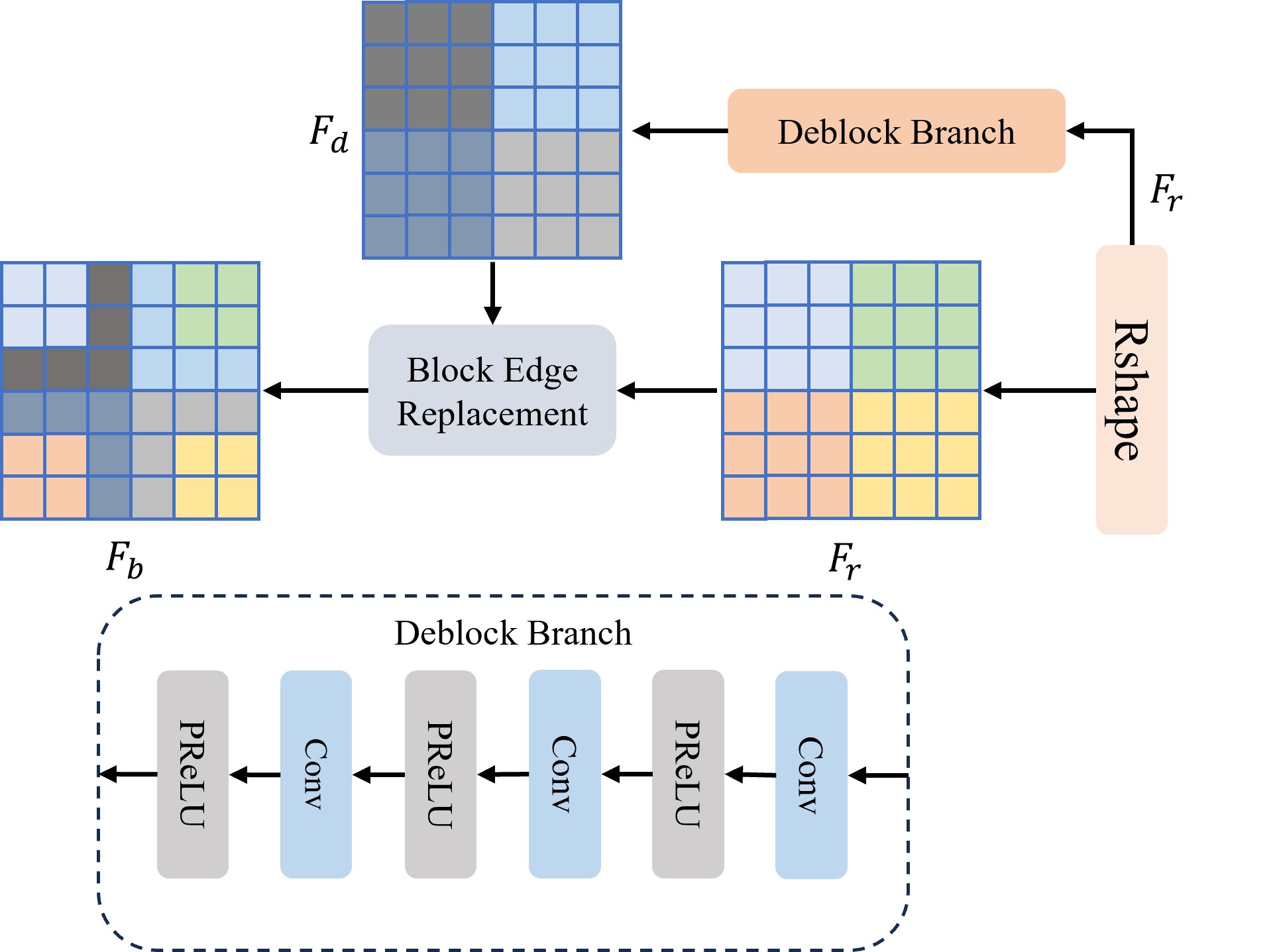}
\end{center}
   \caption{Illustration of the block edge replacement policy and deblock branch structure.}
\label{fig:replace}
\end{figure}

Directly stitching these blocks into a full image feature to process would inevitably cause block artifacts, so we introduce the deblock branch.
As shown in Fig. \ref{fig:replace}, Deblock branch consists of three layers of convolution and PReLU activation function. The features ${F}_{k_1}$, ${F}_{k_2}$, ${F}_{k_3}$ are reshaped into a full feature image ${F}_{r}$ according to their positions. The ${F}_{r}$ passes through the deblock branch to obtain ${F}_{d}$, which is used to replace the block edges in ${F}_{r}$ and resulting in ${F}_{b}$. Finally, an Upsample module outputs the SR image ${I}_{SR}$ .

We train the JointSR model as an integrated unit. Three SRNet Feature Extraction layers use pre-trained weights.

\begin{table*}[ht]
\renewcommand\arraystretch{1.1}
\centering
\begin{tabular}{|c |c |c c |c c |c|} \hline    
\multirow{2}{*}{\textbf{Evaluation index}} & \multirow{2}{*}{\textbf{SR method}} & \multicolumn{2}{|c|}{\textbf{BICUBIC}}& \multicolumn{2}{|c|}{\textbf{Down-Net}}& \multirow{2}{*}{\makebox[0.07\textwidth][c]{\textbf{Mean}}} \\ \cline{3-6}
 &  & \makebox[0.07\textwidth][c]{Test2K} & \makebox[0.07\textwidth][c]{Test4K} & \makebox[0.07\textwidth][c]{Test2K} & \makebox[0.07\textwidth][c]{Test4K} &  \\ \hline 
\multirow{4}{*}{PI↓} & OmniSR-O & 6.3213 & 5.7104 & 5.4829 & 4.8712 & 5.5964 \\ 
 & BBMR-OmniSR & \textbf{5.5386} & \textbf{5.0819} & \textbf{4.6909} & \textbf{4.2477} & \textbf{4.8897} \\ \cline{2-7}
 & CRAFT-O & 6.4482 & 5.8717 & 5.4919 & 4.8938 & 5.6764 \\ 
 & BBMR-CRAFT & \textbf{5.5865} & \textbf{5.1480} & \textbf{4.6519} & \textbf{4.2273} & \textbf{4.9034} \\ \hline 
\multirow{4}{*}{NIQE↓} & OmniSR-O & 6.6214 & 5.9650 & 5.6639 & 5.0877 & 5.8345 \\ 
 & BBMR-OmniSR & \textbf{5.5452} & \textbf{5.0814} & \textbf{4.5704} & \textbf{4.2640} & \textbf{4.8652} \\ \cline{2-7}
 & CRAFT-O & 6.7598 & 6.1989 & 5.6746 & 5.1009 & 5.9335 \\ 
 & BBMR-CRAFT & \textbf{5.5626} & \textbf{5.2774} & \textbf{4.4778} & \textbf{4.2302} & \textbf{4.887}\textbf{0} \\ \hline 
\multirow{4}{*}{PSNR↑} & OmniSR-O & 36.50 & 33.15 & 39.60 & 36.75 & 36.50 \\ 
 & BBMR-OmniSR & \textbf{37.78} & \textbf{34.30} & \textbf{41.36} & \textbf{38.49} & \textbf{37.98} \\ \cline{2-7}
 & CRAFT-O & 36.32 & 32.99 & 39.46 & 36.64 & 36.35 \\ 
 & BBMR-CRAFT & \textbf{37.53} & \textbf{34.05} & \textbf{41.43} & \textbf{38.46} & \textbf{37.86} \\ \hline
 
\end{tabular}
\caption{PI\cite{PI}, NIQE\cite{NIQE} and PSNR values on Test2K, Test4K. The best results are remarked in bold font. O: the original networks with overall image scaling. BBMR: Block-Based Multi-Scale Image Rescaling Framework. OmniSR and CRAFT denote upscaling models. BICUBIC and Down-Net denote downscaling methods.}
\label{tab:total contrast}
\end{table*}

\begin{table*}[ht]
\renewcommand\arraystretch{1.2}
\small
\centering
\begin{tabular}{|c|c|cc|cc|} \hline      
\makebox[0.18\textwidth][c]{\textbf{SR method}} & \makebox[0.1\textwidth][c]{\textbf{Parameters}} & \makebox[0.1\textwidth][c]{\textbf{Test2k}} & \makebox[0.1\textwidth][c]{\textbf{FLOPs}} & \makebox[0.1\textwidth][c]{\textbf{Test4k}} & \makebox[0.1\textwidth][c]{\textbf{FLOPs}} \\ \hline 

OmniSR-O & 0.77M & 36.50 & \textbf{243.46G(100\%)}& 33.15 & \textbf{327.87G(100\%)}\\ 
\makecell{BBMR-OmniSR} & 2.3M & \textbf{37.78} & 250.15G(103\%)& \textbf{34.30} & 337.22G(103\%)\\ \hline 

CRAFT-O & 0.72M & 36.32 & \textbf{242.20G(100\%)}& 32.99 & \textbf{326.17G(100\%)}\\ 
\makecell{BBMR-OmniSR} & 2.20M & \textbf{37.53} & 249.04G(103\%)& \textbf{34.05} & 335.61G(103\%)\\ \hline

\end{tabular}
\vspace{0.2cm}
\caption{Comparison of Parameters, PSNR and FLOPs using bicubic downscaling method and different upscaling models. Parameters and FLOPs only computes the Upscaling Module. The best results are remarked in bold font.}
\label{tab:FLOPs with Upscaling Module}
\end{table*}

\subsection{Very Lightweight Super-Resolution Model}
We find that SR images obtained through deep learning-based super-resolution exhibit certain similarities. There are computational limitations in some cases. To reduce Downscaling Module computation, we propose using a fixed very lightweight super-resolution model instead of dynamically adjusting the Downscaling Module's SR model to match the SR model in Upscaling Module. This work not only significantly reduces the computational load on the server but also can reduce the overhead of model replacement.

To achieve this very lightweight super-resolution model, we streamline the architecture by reducing the number of OSAG modules from 5 to 1, based on the OmniSR \cite{wang2023omni}. This reduction maintains the model's ability to effectively upscale images while dramatically cutting down the computational requirements.


\section{Experiment}
\subsection{Setting}
Our proposed BBMR is a general image rescaling framework that can be applied to various image rescaling methods. For verification, we selected OmniSR \cite{wang2023omni} and CRAFT \cite{CRAFT} as super-resolution models and employed both BICUBIC downscaling and a neural network-based downscaling method proposed by \cite{global}. This allows us to effectively demonstrate the effect of our framework. For ease of experimentation, we select ×2, ×4, and ×8 as our multi-scale factors, with the overall scaling factor set to ×4.

\subsubsection{Training Data}
We use the DIV2k and Flickr2k \cite{div2k_flickr} dataset for training. With BICUBIC downscaling, LR image sizes are 64×64 for ×2 and ×4 scaling factors, and 32×32 for ×8. Down-Net employs end-to-end training with 256×256 HR input and SR output. For JointSR, we randomly crop 768×768 regions from DIV2k and Flickr2k as HR images and divide them into 36 HR sub-blocks with the size of 128×128. Subsequently, we use three different scaling rates to randomly select 12 sub-blocks for downscaling as the input of JointSR. Data augmentation includes random horizontal flips and 90/270 degree rotations.

\subsubsection{Testing data}
Test2K and Test4K each contain 100 images, which were selected from the DIV8K \cite{gu2019div8k} dataset and downscaled using bicubic interpolation. PSNR comparisons are performed on the Y channel in YCbCr.

\begin{figure*}[ht]
    \centering
        \includegraphics[width=\textwidth]{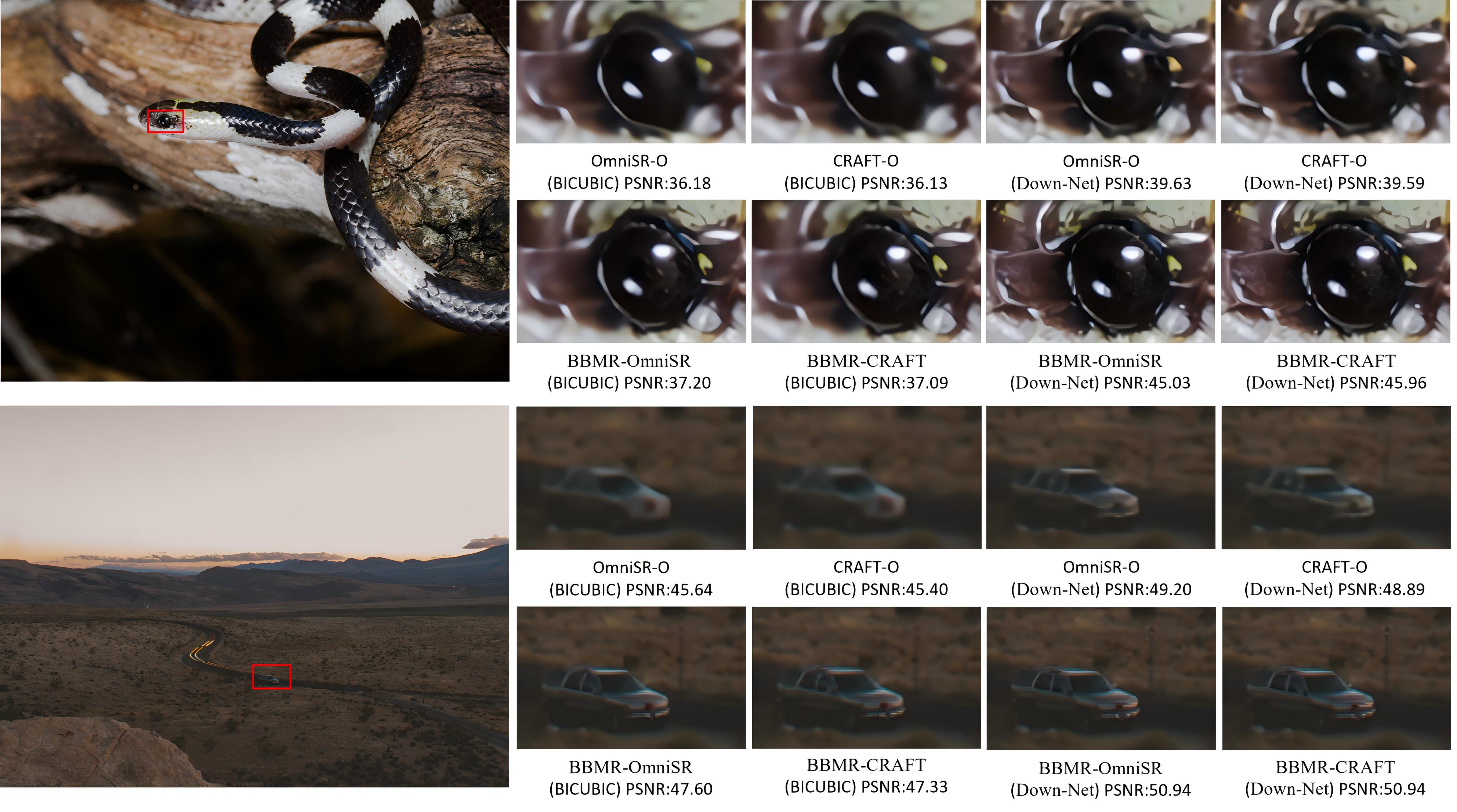}
    \caption{Comparison of visual quality between the BBMR method and traditional image rescaling method using different downscaling and upscaling approachs.}
\label{fig:visual qualioty}
\end{figure*}

\begin{table*}[h]
\centering
\begin{tabular}{|c |c c |c c|} \hline      
\makebox[0.25\textwidth][c]{\textbf{SR method}} & \makebox[0.1\textwidth][c]{\textbf{Test2k}} & \makebox[0.1\textwidth][c]{\textbf{FLOPs}} & \makebox[0.1\textwidth][c]{\textbf{Test4k}} & \makebox[0.1\textwidth][c]{\textbf{FLOPs}} \\ \hline 
BBMR-OmniSR

w/o light & \textbf{37.78} & 1258.53G(100\%) & \textbf{34.30} & 1694.85G(100\%) \\ 
BBMR-OmniSR

w/ light & 37.57 & \textbf{324.01G}\textbf{(26\%)} & 34.11 & \textbf{436.34}\textbf{(26\%)} \\ \hline 
BBMR-CRAFT

w/o light & \textbf{37.53} & 1256.80G(100\%) & \textbf{34.05} & 1692.52G(100\%) \\ 
BBMR-CRAFT

w/ light & 37.39 & \textbf{324.01G}\textbf{(26\%)} & 33.93 & \textbf{436.34G}\textbf{(26\%)} \\ \hline
\end{tabular}
\caption{Comparison of the computational load of the Downscaling Module using a very lightweight super-resolution model and the original super-resolution model. FLOPs only computes the Downscaling Module. The best results are remarked in bold font.}
\label{tab:FLOPs with Downscaling Module}
\end{table*}

\subsubsection{Training details}
The batch size for OmniSR and CRAFT at all upscaling rates is set to 32, trained for 1000 epochs. JointSR uses a batch size of 4 and trained for 100 epochs. All networks start with a learning rate of 0.0005, for SR networks halved every 250 epochs, and for JointSR halved every 20 epochs. Training utilizes the AdamW optimizer with L1 loss. All models are built with PyTorch and trained on NVIDIA GeForce RTX 4090 GPUs. The loss of end-to-end training is shown below, where k is the scaling factor.

\begin{equation}
\small
{Loss}_{p2p}=(\frac{1}{2k}){ L1Loss(LR', LR)}+{ L1Loss(SR, HR)}\label{loss1}
\end{equation}

$LR'$ represents the output obtained through Down-Net downscaling, while $LR$ is derived via bicubic downscaling. $SR$ is the output after super-resolution, and $HR$ is the original high-resolution image.

\subsection{Evaluation Index Of BBMR}
We select PSNR, NIQE, and PI as the evaluation metrics for our experiments, where NIQE and PI are no-reference visual evaluation metrics. As shown in Tab. \ref{tab:total contrast}, no matter which image rescaling method is chosen, our BBMR framework demonstrates significant improvements on the Test2k and Test4k datasets compared to the original network method. The PSNR value is improved by approximately 1.5dB on average. Notably, when using the neural network downscaling method, the PSNR value can be enhanced by up to 1.97dB compared to the original network method. Furthermore, the substantial improvement in NIQE and PI scores suggests that our BBMR framework can enhance the overall visual quality of image super-resolution.

\subsection{Visual Quality Comparison Of BBMR}
Fig. \ref{fig:visual qualioty} shows the visual quality comparison of the BBMR framework using different image rescaling methods compared to the original network methods. It is evident that for primary subjects like cars or animals, BBMR assigns them a lower scaling rate, resulting in much clearer images.

\begin{figure*}[ht]
    \centering
    \includegraphics[width=1\textwidth]{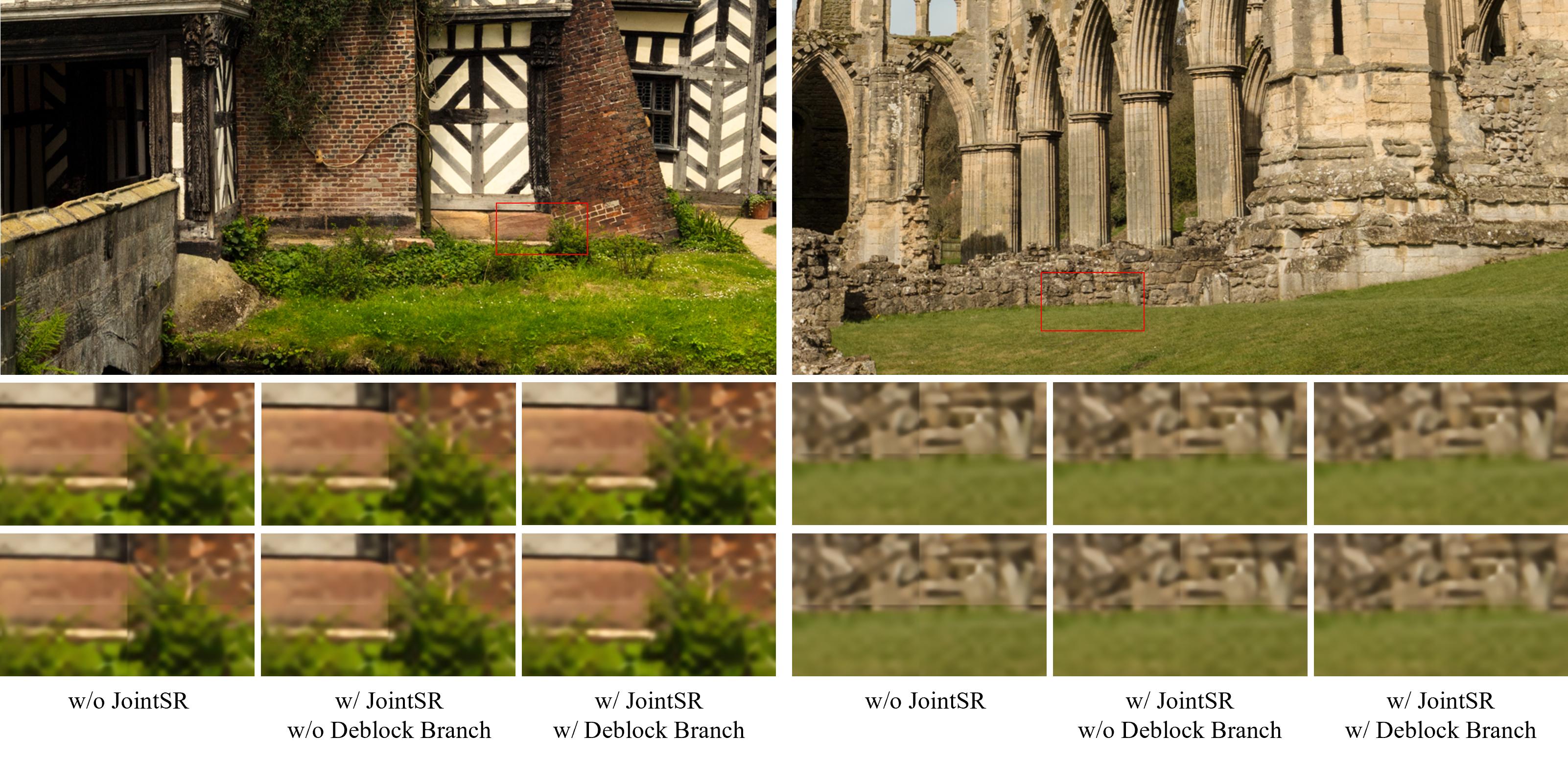}
    \caption{Visual quality comparison of block effects using bicubic downscaling. The boundary of the image block is located at the cross in the middle of the image. The first row shows the upscaling results of OmniSR, while the second row shows the upscaling results of CRAFT.}
\label{fig:deblock}
\end{figure*}

\subsection{Comparison Of The Calculation Amount Of Upscaling Module}
As shown in Tab. \ref{tab:FLOPs with Upscaling Module}, our BBMR method achieves a significant improvement in quality while only increasing the computational load in the Upscaling Module by 3\% compared to the original network method, which is almost negligible. BBMR ensures that the overall scaling rate size of the LR image remains unchanged, which indicates that our strategy can significantly enhance image super-resolution quality while keeping the transmission data size and the computational load in the Upscaling Module almost constant.

\subsection{The Effect Of Very Lightweight Super-Resolution Model}
We used the very lightweight super-resolution models to assist in allocating the scaling rate of image sub-blocks instead of using the super-resolution model employed on the Upscaling Module. As shown in Tab. \ref{tab:FLOPs with Downscaling Module}, using a very lightweight super-resolution model in the Downscaling Module reduces the computation by approximately three-quarters compared to the previous method, with a PSNR decrease ranging from 0.21db to 0.12db. This is very useful under conditions of limited server-side computational capacity.

\subsection{Ablation Study}
\vspace{-0.2cm}

\begin{table}[ht]
\centering
\begin{tabular}{|c |c c c c |c |c|} \hline     
\textbf{Case} & \makebox[0.03\textwidth][c]{\textbf{Block}} & \makebox[0.03\textwidth][c]{\textbf{BMD}} & \makebox[0.03\textwidth][c]{\textbf{Joint}} & \makebox[0.03\textwidth][c]{\textbf{DB}} & \textbf{PSNR} & \textbf{FLOPs} \\ \hline 
1 & × & × & × & × & 33.15 & 327.87G \\ 
2 & $\surd$ & × & × & × & 33.08 & 327.65G \\  
3 & $\surd$ & $\surd$ & × & × & 34.26 & 328.30G \\  
4 & $\surd$ & $\surd$ & $\surd$ & × & 34.32 & 330.53G \\  
5 & $\surd$ & $\surd$ & $\surd$ & $\surd$ & 34.30 & 337.22G \\ \hline

\end{tabular}
\caption{Ablation study of the proposed BBMR on Test4K dataset with OmniSR upscaling and BICUBIC downscaling for 4× SR. Block: image super-resolution with average block. BMD: using the block-based multi-scale downscaling strategy. Joint: Using the JointSR method. DB: Using the JointSR method with deblock branch. FLOPs only computes the Upscaling Module.}
\label{tab:ablation study}
\end{table}

\subsubsection{The role of block-based multi-scale downscaling strategy}
As shown in Tab. \ref{tab:ablation study}, simply dividing the image into blocks and using the same scaling rate will degrade the quality of the super-resolved image. This is because each image block loses information at the edges and creates severe block artifacts. In contrast, using our block-based multi-scale downscaling strategy shows significant improvement on the Test4k dataset, indicating that our block-based multi-scale downscaling strategy can reduce the quality loss caused by blocking and enhance the overall quality of super-resolved images.

\subsubsection{The role of deblock branch In JointSR}
As shown in Tab. \ref{tab:ablation study}, although the PSNR of the JointSR method w/o deblock branch is 0.02dB higher on Test4k compared to w/ deblock branch, the block artifacts w/o deblock branch are very obvious, as seen in Fig. \ref{fig:deblock}. In contrast, the block artifacts are almost completely eliminated after adding the deblock branch.

\subsubsection{The role of JointSR}
JointSR is our proposed method that integrates super-resolution with deblocking at the feature level to reduce computational cost. As shown in Tab. \ref{tab:ablation study}, the JointSR w/ deblock branch increases the PSNR by 0.04db on the Test4k dataset and JointSR w/o deblock branch increases the PSNR by 0.06db with almost unchanged computation. Furthermore, as illustrated in Fig. \ref{fig:deblock}, it is evident that JointSR can effectively remove blocking artifacts, no matter which SR model is used.

\section{Conclusion}

This paper proposes a Block-Based Multi-Scale Image Rescaling (BBMR) framework. Additionally, within the Upscaling Module, we propose a joint super-resolution method (JointSR) to remove image blocking artifacts. The key idea is to selectively allocate the scaling rate for each image block based on the difficulty of its super-resolution, while keeping the overall scaling rate constant. Extensive experiments show that the framework improves the PSNR by 1.06 to 1.96 dB on the Test2K and Test4K test sets.

\section{Acknowledgments}
This work was supported in part by the National Science Foundation of China (62172153) and Hunan Provincial Key Research and Development Program of China (2024AQ2032).


\bibliography{Formatting-Instructions-LaTeX-2025}

\end{document}